\documentclass[twocolumn,english]{revtex4}
\usepackage[T1]{fontenc}
\usepackage[latin9]{inputenc}
\usepackage{float}
\usepackage{textcomp}
\usepackage{amsmath}
\usepackage{graphicx}

\makeatletter

\providecommand{\tabularnewline}{\\}

\makeatother

\usepackage{babel}

\begin{document}

\title{Lens design based on instantaneous focal function}

\author{Sung Nae Cho}

\email{sungnae.cho@samsung.com}

\affiliation{MEMS \& Packaging Group, Micro Systems Lab, Samsung Advanced Institute
of Technology, Mt. 14-1 Nongseo-dong Giheung-gu, Yongin-si Gyeonggi-do,
446-712, South Korea.}

\date{Prepared on January 30, 2009}

\begin{abstract}
The formula for the lens is derived based on the information of instantaneous
focal function. Focal function is an important tool in designing lenses
with extended depth of focus (EDoF) because this allows \textbf{EDoF}
lens designers to try out various mathematical curves using computers
to optimize their design. Once an optimal focal function information
is obtained, the corresponding physical \textbf{EDoF} lens can be
fabricated using the lens equation formulated in this presentation.
\end{abstract}
\maketitle

\section{Introduction}

\begin{figure}[H]
\begin{centering}
\includegraphics[scale=0.6]{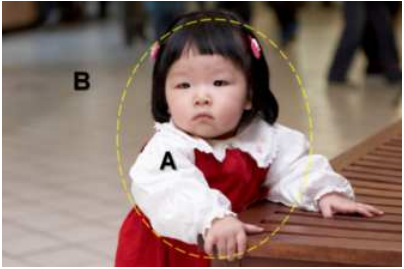}
\par\end{centering}

\caption{\label{fig:DOF} Illustration of depth of field. Region \textbf{A}
is within the depth of field (\textbf{DOF}), whereas the region \textbf{B}
is outside of \textbf{DOF}. The image is blurred drastically in region
\textbf{B}.}

\end{figure}

\begin{figure}[H]
\begin{centering}
\includegraphics[scale=0.5]{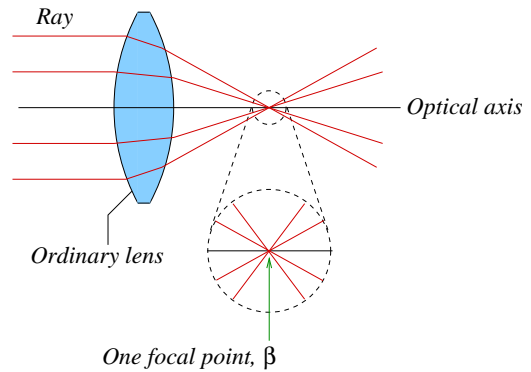}
\par\end{centering}

\caption{\label{fig:ordinary_lens} An ideal ordinary lens is characterized
by a single focal point.}

\end{figure}

Optical imaging system with large depth of field (\textbf{DOF}) is
required to produce sharp images\citet{DOF_wikipedia_site}. In photography,
the \textbf{DOF} is the portion of a scene that appears sharp in the
image, for example, the region denoted by \textbf{A} in Fig. \ref{fig:DOF}.
Ordinarily, a lens focuses parallel rays of light at one distance
known as the focal point, as illustrated in Fig. \ref{fig:ordinary_lens}.
Therefore, not all points within the \textbf{DOF} can be claimed as
focused per se. However, due to a gradual decrease in the sharpness
of the image from the focused spot, the amount of blurring within
the \textbf{DOF} is imperceptible to human eyes under normal viewing
conditions. As such, in particularly for films and photography, the
image region can be subdivided into two, where one lies within and
the other lies external to the \textbf{DOF}. In the photograph of
Fig. \ref{fig:DOF}, the region \textbf{A}, wherein the image appears
sharp and focused, is said to lie within the \textbf{DOF}; whereas
the region \textbf{B}, in which region the image is blurred, is said
to lie external to the \textbf{DOF}. 

Alternatively, but equivalently, the \textbf{DOF} in an imaging system
is defined as the distance in the object space in which objects are
considered to be in focus. The distance over which objects appear
sharp can be increased by extending the \textbf{DOF} of an imaging
system. Traditionally, the \textbf{DOF} of an imaging system can be
increased by either decreasing the size of lens aperture or by increasing
the shutter speed, or through tweaking of the both. These methods,
however, drastically reduces the amount of light passing through the
lens and require extra lighting. For developing still images, the
required extra lighting can be accommodated by the use of a flash.
For motion pictures, however, this approach proves to be inadequate,
as typical video involves about thirty frames of images per   second.%
\footnote{According to the industry standard of specifications set by the flat
panel display consortium, the \textbf{1080P} specification of the
full-\textbf{HD} quality of liquid crystal displays (\textbf{LCD}s)
process sixty image frames per second.%
} In addition to this difficulty, the smaller lens aperture increases
diffraction and this places a practical limit on the extent to which
the \textbf{DOF} of an optical imaging system can be enhanced by the
aforementioned methods. That being said, can the \textbf{DOF} of an
imaging system be increased (A) without sacrificing the intensity
of light passing through the lens and (B) without increasing the diffraction?
The answer to this question is yes and this approach involves some
sort of digital filtering. 

The digital filtering method requires a scheme for the image reconstruction
algorithm based on the principles of wave optics\citet{Portney,Mendlovic,Ben-Eliezer,Dowski,Bradburn,Bradburn2}.
The image reconstruction code is often hardcoded in the accompanying
digital processing unit (\textbf{DPU}) and, for this reason, the imaging
method based on digital filtering is coined as the software assisted
imaging technology or \textbf{SAIT} for short.%
\footnote{Unfortunately, the work place I am affiliated with, Samsung Advanced
Institute of Technology, also uses the abbreviation {}``SAIT'' for
the name. To distinguish this SAIT from software assisted imaging
technology, I denote the latter with bold faced version, \textbf{SAIT}. %
} The \textbf{SAIT} solution for an imaging system is a promising technology
in that it has potential to increase the \textbf{DOF} while encompassing
altogether the processes of which (A) drastically reduce the amount
of light passing through the lens and (B) those problems associated
with increased diffraction due to the reduced diameter of the  lens
aperture. 

The software assisted imaging technology is not as perfect as it sounds
and it has some of its own problems to offer. Among them, one directly
relates to the image processing speed. In the traditional imaging
solutions based on lenses, which I refer to as the analog technology,
complex arrangement of lenses function to focus image at the focal
plane from wherein the image is developed. As such systems based on
analog technology by-pass the digital filtering stage altogether,
images are developed instantaneously in analog imaging systems. However,
due to the aforementioned problems and limitations of the traditional
method for increasing the \textbf{DOF}, the analog technology provides
only a limited solution when concerning the image quality. The systems
based on \textbf{SAIT} principle face problems that are exactly the
opposite in nature from that of the analog systems. In principle,
the reconstructed image by a way of image reconstruction algorithm
can be made to resemble the original image to any level or degree
of resemblance, provided the image processing speed is of no concern.
But, such methodology would limit systems based on \textbf{SAIT} to
still imagery applications and the video sector of the market must
be discarded, which is a bad idea for business. Naturally, for systems
based on \textbf{SAIT} principle,  a trade off must be made between
the image quality and the image processing speed. 

\begin{figure}[H]
\begin{centering}
\includegraphics[scale=0.5]{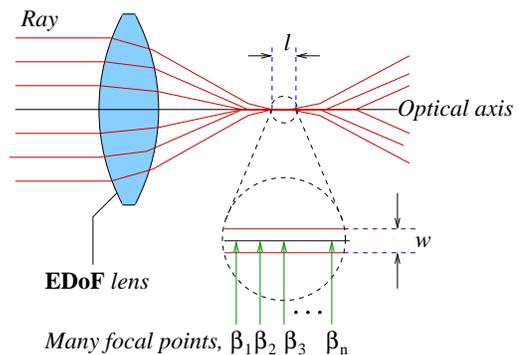}
\par\end{centering}

\caption{\label{fig:EDoF_lens} Lens with extended depth of focus has many
focal points, $f_{1},$ $f_{2},$ $f_{3},$ and so on. An ideal \textbf{EDoF}
lens has infinitely many focal points and all light rays are confined
within the cylindrical tube very small diameter $w$ and very long
length $l.$ }

\end{figure}

Among the early pioneers to successfully commercialize imaging system
based on \textbf{SAIT} are Cathay and Dowski, who did much of their
work at the University of Colorado\citet{Dowski,Bradburn,Bradburn2}.
In the modern literature, their work is cited as {}``wavefront coding.''
The \textbf{SAIT} solution based on wavefront coding has been trade
marked by CDM Optics, Inc., and it is known as the $\textnormal{Wavefront }\textnormal{Coding}^{\textup{TM}}.$
The imaging solution based on wavefront coding basically involves
two stages: the input and the output stages. The input stage involves
 the optical element and this represents the hardware contribution
side of the \textbf{SAIT} solution. The output stage involves the
\textbf{DPU}, wherein the image reconstruction code base is hardcoded,
and this represents the software contribution side of the \textbf{SAIT}
solution. The optical element in \textbf{SAIT} solution is distinguished
from the input stage of traditional imaging system, which is just
complex series of ordinary lenses, in that it produces many focal
points along the optical axis instead of just one at the focal plane.
Such optical element is referred to as lens having extended depth
of focus (or \textbf{EDoF} for short) and this is illustrated in Fig.
\ref{fig:EDoF_lens}. 

Both the quality of reconstructed image and the image reconstruction
speed are critically important in \textbf{SAIT}. In principle, the
ordinary lens, such as the one illustrated in Fig. \ref{fig:ordinary_lens},
can just as well serve as the input stage for the \textbf{SAIT} imaging
systems. However, this must be done at the cost of overly complicated
algorithm routines for the software side of the system. The length
of image reconstruction code directly relates to the number of transistors
in a \textbf{DPU}. As a simple rule of a thumb, more transistors there
are in a \textbf{DPU}, more energy it requires to operate. And, more
lines of coding for the image reconstruction algorithm implies the
slower processing speed for the reconstructed image output. The full-\textbf{HD}
quality of a video involves sixty image frames every second. This
implies, the application of SAIT system to full-HD video processing
would require the image reconstruction time span of $16\,\textup{ms}$
or less. The demand for very fast processing speed and low power consumption
make ordinary lenses inadequate for the input stage of the \textbf{SAIT}
system, which leave open for an alternate solution for the optical
element to be used as the input stage. 

The image reconstruction code base can be optimized for image processing
speed and quality if the \textbf{EDoF} lens is used as the optical
element for the input stage of \textbf{SAIT} system. The \textbf{EDoF}
lens is characterized by parameters $l$ and $w.$ The parameter $l$
represents the depth of focal points along the optical axis and the
parameter $w$ represents the width of bundle containing light rays
associated with each focal points, as illustrated in Fig. \ref{fig:EDoF_lens}.
The ideal EDoF lens has the parameter $l$ of which is infinite in
length and the parameter $w$ of which is infinitely thin. For the
realistic \textbf{EDoF} lenses, however, the parameters $l$ and $w$
are typically on the order of microns.  

\begin{figure}[H]
\begin{centering}
\includegraphics[scale=0.5]{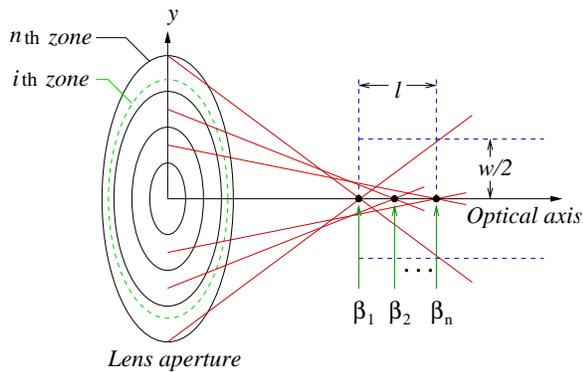}
\par\end{centering}

\caption{\label{fig:Getman_EDoF_lens} Schematic of conceptual \textbf{EDoF}
lens proposed by Alexander and Lukyanov. The lens aperture has an
axial symmetry about the optical axis.}

\end{figure}

\begin{figure}[H]
\begin{centering}
\includegraphics[scale=0.65]{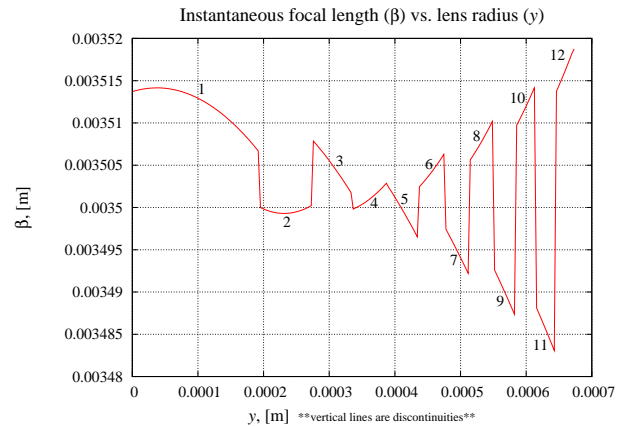}
\par\end{centering}

\caption{\label{fig:lens-data-Getman} Focal function $\beta$ proposed by
Alexander and Lukyanov \citet{Getman}. }

\end{figure}

The idea of \textbf{EDoF} lens as an optical element which focuses
light into longitudinally directed line along the optical axis was
first proposed by Golub, et. al. \citet{Golub}. The idea was later
adopted by others and it has found applications in various imaging
systems, such as microscopes, cameras, and lithography to list a few,
\citet{Portney,Mendlovic,Ben-Eliezer,Dowski,Forster,Liu,Getman}.
Alexander and Lukyanov have recently proposed a conceptual scheme
for the \textbf{EDoF} lens\citet{Getman}. Their scheme for \textbf{EDoF}
lens consists of zones that are axially symmetric about the optical
axis and this is illustrated in Fig. \ref{fig:Getman_EDoF_lens}.
The idea behind their concept is as follows. The light ray crossing
each zone gets focused to a unique spot on the optical axis and this
spot is within in the \textbf{EDoF} lens parameter $l.$ If $\beta_{i}$
is the function which describes the focal length for the $i\,\textup{th}$
concentric zone in Fig. \ref{fig:Getman_EDoF_lens}, the lens aperture,
in principle, can be tailored to behave like \textbf{EDoF} lens by
tweaking $\beta_{i}.$ Borrowing their terminology, the $\beta_{i}$
is referred to as the instantaneous focal function. The idea behind
\textbf{EDoF} lens is to make $l$ as large and $w$ as small as possible.
By experimenting with different functions for $\beta_{i},$ the \textbf{EDoF}
lens parameters $l$ and $w$ can be engineered to the acceptable
range for the imaging systems based on \textbf{SAIT} principle. This
is exactly just what Alexander and Lukyanov did, and their focal function
is summarized in Fig. \ref{fig:lens-data-Getman}. In the figure,
the vertical lines represent discontinuities and each of the twelve
zones has been indicated appropriately by a number. For each of the
zones in Fig. \ref{fig:Getman_EDoF_lens}, the incident parallel rays
are focused at different points on the optical axis within $l;$ and,
 the focusing is described by the instantaneous focal function, $\beta=\beta_{i},$
summarized in Fig. \ref{fig:lens-data-Getman}. 

The imaging system based on \textbf{SAIT} is a package solution in
which both the hardware (optical element or \textbf{EDoF} lens) and
the software (image reconstruction code base) contributions must be
optimized. Alexander and Lukyanov experimented with various mathematical
curves for the focal function $\beta$ in an attempt to make the \textbf{EDoF}
lens parameters $l$ and $w$ as ideal as possible, but did not attempt
to provide any solutions concerning the physical shape or the profile
for their conceptual \textbf{EDoF} lens. To test the image reconstruction
code for the processing speed and the quality of generated image output,
the information on the point spread function (or \textbf{PSF} for
short) of the optical element input stage is required. Since no information
on the physical profile of their conceptual \textbf{EDoF} lens was
available, the test for their image reconstruction algorithm had to
be deferred. It was my job to design a physical \textbf{EDoF} lens.
Since Alexander's image reconstruction code was based on the input
from a lens aperture satisfying the focal function described in Fig.
\ref{fig:lens-data-Getman}, the physical \textbf{EDoF} lens to be
designed had a constraint of satisfying the same focal function characteristics.
To end the story, the physical \textbf{EDoF} lens with such characteristics
for the focusing behavior was found\citet{Cho}. The obtained physical
profile of lens was entered into CODE V\textregistered{}%
\footnote{CODE V\textregistered{} is an optical design program with graphical
user interface for image forming and fiber optical systems by Optical
Research Associates (ORA), an organization that has been supporting
customer success for over 40 years, \underbar{www.opticalres.com}.%
} to generate the needed \textbf{PSF} information for the lens aperture.%
\footnote{Using CODE V\textregistered{} to obtain \textbf{PSF} information turned
out to be a nontrivial task, as the built-in curve fitting functions
for CODE V\textregistered{} were not too {}``happy'' with the obtained
result for the surface profile of \textbf{EDoF} lens.%
} This information was in turn used by Alexander to test for the performance
of his algorithm. 

This work concerns the result of my role in the project, which was
to design a physical \textbf{EDoF} lens from the focal function characteristics.
As such, this work concerns the hardware contribution side of the
optical imaging system based on \textbf{SAIT}.

\section{Theory}

\subsection{Axial symmetry}

\begin{figure}[H]
\begin{centering}
\includegraphics[scale=0.5]{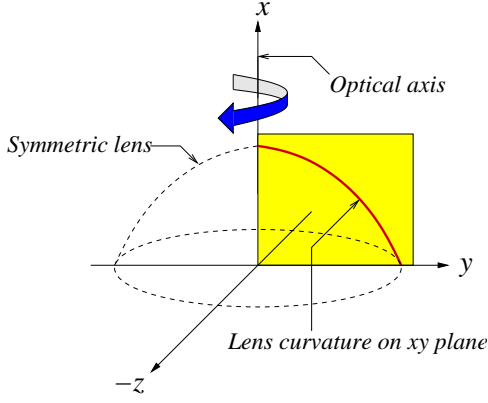}
\par\end{centering}

\caption{\label{fig:3D_sphere} Schematic of lens symmetric about optical axis. }

\end{figure}

By definition, an axial symmetry is a symmetry about a given given
axis. The object has an axial symmetry if its appearance is unchanged
with the rotation about some axis. Illustrated in Fig. \ref{fig:3D_sphere}
is a schematic of conceptual lens, which shows an axial symmetry about
the optical axis, i.e., the $x$ axis. Such a lens can be dissected
through the origin with the $xy$ plane as shown in Fig. \ref{fig:3D_sphere}.
The curve traced on the $xy$ plane, which is a set of points on the
surface of lens, can be revolved about the optical axis for the three
dimensional shape of the lens. The design of axially symmetric lens,
therefore, simplifies to the problem of finding the set of points
on the surface of lens of which gets traced on the $xy$plane. 

The physical law of which governs the bending of light is the Snell's
law. I shall apply Snell's principle to derive the equation for the
cross-sectional profile of the lens. The terminology, {}``cross-sectional
profile of the lens,'' implies the lens curvature on the $xy$ plane,
which is illustrated in Fig. \ref{fig:3D_sphere}.

\subsection{Derivation from Snell's law}

\begin{figure}[H]
\begin{centering}
\includegraphics[scale=0.5]{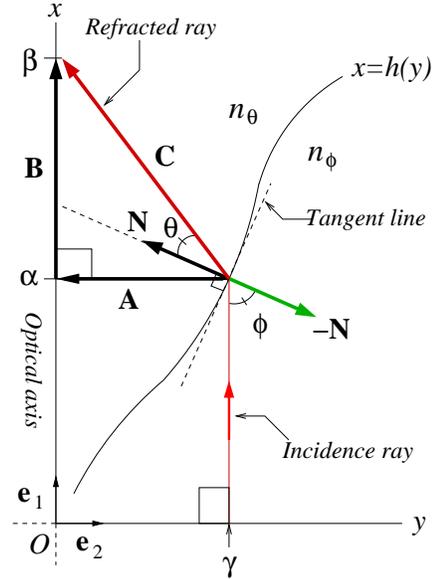}
\par\end{centering}

\caption{\label{fig:Law_of_refraction} Illustration of Snell's law. The scalar
quantity $\beta$ denotes the focal length. }

\end{figure}

When a ray of light passes across media of different refractive indices,
its path is governed by the Snell's law, \begin{equation}
n_{\phi}\sin\phi=n_{\theta}\sin\theta,\label{eq:Snell-law}\end{equation}
 as illustrated in Fig.  \ref{fig:Law_of_refraction}. Here, $n_{\phi}\equiv n_{\phi}\left(\omega\right)$
and $n_{\theta}\equiv n_{\theta}\left(\omega\right)$ are frequency
dependent refractive indices with $\omega$ denoting the angular frequency
of the light. The parameters $\phi$ and $\theta$ represent the angle
of incidence and angle of refraction, respectively.  

If $\mathbf{N}$ denotes the normal vector to the local point $y=\gamma$
on the curve $x=h\left(y\right),$ then it can be shown \[
\left\Vert -\mathbf{N}\times\left(-\mathbf{e}_{1}\right)\right\Vert =\left\Vert -\mathbf{N}\right\Vert \left\Vert -\mathbf{e}_{1}\right\Vert \sin\phi=N\sin\phi\]
 and the expression for $\sin\phi$ becomes \begin{eqnarray}
\sin\phi=\frac{\left\Vert \mathbf{N}\times\mathbf{e}_{1}\right\Vert }{N}, &  & N\equiv\left\Vert \mathbf{N}\right\Vert ,\label{eq:sin-of-phi}\end{eqnarray}
 where $\mathbf{e}_{1}$ is the unit basis for the $x$ axis. 

Similarly, the expression for $\sin\theta$ may be obtained by considering
vectors $\mathbf{A},$ $\mathbf{B},$ and $\mathbf{C}$ of Fig.  \ref{fig:Law_of_refraction}.
The vectors $\mathbf{A},$ $\mathbf{B},$ and $\mathbf{C}$ satisfy
the relation, \begin{equation}
\mathbf{A}+\mathbf{B}=\mathbf{C}.\label{eq:C_pre1}\end{equation}
 In explicit form, vectors $\mathbf{A}$ and $\mathbf{B}$ are defined
as \begin{equation}
\mathbf{A}=-\gamma\mathbf{e}_{2},\quad\mathbf{B}=\left(\beta-\alpha\right)\mathbf{e}_{1},\label{eq:C_pre2}\end{equation}
 where $\mathbf{e}_{2}$ is the unit basis for the $y$ axis. With
Eqs. (\ref{eq:C_pre1}) and (\ref{eq:C_pre2}), the vector $\mathbf{C}$
becomes \begin{equation}
\mathbf{C}=\left(\beta-\alpha\right)\mathbf{e}_{1}-\gamma\mathbf{e}_{2}.\label{eq:C}\end{equation}
 The vector cross product $\mathbf{N}\times\mathbf{C}$ is given by
\[
\mathbf{N}\times\mathbf{C}=\left(\beta-\alpha\right)\mathbf{N}\times\mathbf{e}_{1}-\gamma\mathbf{N}\times\mathbf{e}_{2}\]
 and its magnitude becomes \begin{align}
\left\Vert \mathbf{N}\times\mathbf{C}\right\Vert  & =\left\Vert \left(\beta-\alpha\right)\mathbf{N}\times\mathbf{e}_{1}-\gamma\mathbf{N}\times\mathbf{e}_{2}\right\Vert \nonumber \\
 & =NC\sin\theta,\label{eq:sin-of-theta-pre}\end{align}
 where $N\equiv\left\Vert \mathbf{N}\right\Vert $ and $C=\left\Vert \mathbf{C}\right\Vert .$
Utilizing Eq. (\ref{eq:C}), $C$ may be expressed as  \[
C=\left(\mathbf{C}\cdot\mathbf{C}\right)^{1/2}=\left[\left(\beta-\alpha\right)^{2}+\gamma^{2}\right]^{1/2}\]
 and the Eq. (\ref{eq:sin-of-theta-pre}) is solved for $\sin\theta$
to yield \begin{equation}
\sin\theta=\frac{\left\Vert \left(\beta-\alpha\right)\mathbf{N}\times\mathbf{e}_{1}-\gamma\mathbf{N}\times\mathbf{e}_{2}\right\Vert }{N\left[\left(\beta-\alpha\right)^{2}+\gamma^{2}\right]^{1/2}}.\label{eq:sin-of-theta}\end{equation}

Insertion of Eqs. (\ref{eq:sin-of-phi}) and (\ref{eq:sin-of-theta})
into the Snell's law of Eq. (\ref{eq:Snell-law}) gives \begin{equation}
\frac{n_{\phi}}{n_{\theta}}=\frac{\left\Vert \left(\beta-\alpha\right)\mathbf{N}\times\mathbf{e}_{1}-\gamma\mathbf{N}\times\mathbf{e}_{2}\right\Vert }{\left\Vert \mathbf{N}\times\mathbf{e}_{1}\right\Vert \left[\left(\beta-\alpha\right)^{2}+\gamma^{2}\right]^{1/2}}.\label{eq:Snell-alt-1}\end{equation}
 By definition, the normal vector $\mathbf{N}$ satisfies the relation,
 \[
g\left(x,y\right)=x-h\left(y\right),\]
 where $g\left(x,y\right)$ is a function whose gradient gives $\mathbf{N},$
\[
\mathbf{N}=\nabla g=\frac{\partial g}{\partial x}\mathbf{e}_{1}+\frac{\partial g}{\partial y}\mathbf{e}_{2}=\mathbf{e}_{1}-\frac{\partial h}{\partial y}\mathbf{e}_{2}.\]
 Because $\mathbf{N}$ is the normal vector at the location $\left(x=\alpha,y=\gamma\right),$
I write \begin{equation}
\mathbf{N}=\mathbf{e}_{1}-\left.\frac{\partial h}{\partial y}\right|_{y=\gamma}\mathbf{e}_{2}.\label{eq:N-explicit-xa}\end{equation}
 The following vector cross products are valid, \begin{align*}
\mathbf{N}\times\mathbf{e}_{1} & =\mathbf{e}_{1}\times\mathbf{e}_{1}-\left.\frac{\partial h}{\partial y}\right|_{y=\gamma}\mathbf{e}_{2}\times\mathbf{e}_{1},\\
\mathbf{N}\times\mathbf{e}_{2} & =\mathbf{e}_{1}\times\mathbf{e}_{2}-\left.\frac{\partial h}{\partial y}\right|_{y=\gamma}\mathbf{e}_{2}\times\mathbf{e}_{2},\end{align*}
 where Eq. (\ref{eq:N-explicit-xa}) was used to replace $\mathbf{N}.$
Since $\mathbf{e}_{1}\times\mathbf{e}_{1}=\mathbf{e}_{2}\times\mathbf{e}_{2}=0,$
the previous relations reduce to \begin{equation}
\mathbf{N}\times\mathbf{e}_{1}=\left.\frac{\partial h}{\partial y}\right|_{y=\gamma}\mathbf{e}_{3},\quad\mathbf{N}\times\mathbf{e}_{2}=\mathbf{e}_{3},\label{eq:N-cross-ei}\end{equation}
 where $\mathbf{e}_{3}$ is the unit basis for the $z$ axis of which
satisfies the relation,  \[
\mathbf{e}_{1}\times\mathbf{e}_{2}=\mathbf{e}_{3},\quad\mathbf{e}_{2}\times\mathbf{e}_{1}=-\mathbf{e}_{3}.\]
 Insertion of Eq. (\ref{eq:N-cross-ei}) into Eq. (\ref{eq:Snell-alt-1})
gives \[
\frac{n_{\phi}}{n_{\theta}}=\frac{\left(\beta-\alpha\right)\left.\frac{\partial h}{\partial y}\right|_{y=\gamma}-\gamma}{\left.\frac{\partial h}{\partial y}\right|_{y=\gamma}\left[\left(\beta-\alpha\right)^{2}+\gamma^{2}\right]^{1/2}},\]
 which expression can be rearranged to yield \begin{equation}
\left.\frac{\partial h}{\partial y}\right|_{y=\gamma}=\frac{\gamma}{\beta-\alpha-\frac{n_{\phi}}{n_{\theta}}\left[\left(\beta-\alpha\right)^{2}+\gamma^{2}\right]^{1/2}},\label{eq:Snell-alt-2}\end{equation}
 where $\alpha$ and $\gamma$ are constants of which are depicted
in Fig. \ref{fig:Law_of_refraction}. 

For Alexander and Lukyanov's optical element, the instantaneous focal
function $\beta\equiv\beta\left(y\right)$ in Eq. (\ref{eq:Snell-alt-2})
is as defined in Fig. \ref{fig:lens-data-Getman}. The $\gamma$ for
the $y$ axis is not anything special, of course. Any $y$ belonging
to the domain of $h$ satisfies the Eq. (\ref{eq:Snell-alt-2}). The
generalization of Eq. (\ref{eq:Snell-alt-2}) for all $y$ belonging
to the domain of $h$ is done by making the following replacements:
\[
\alpha\rightarrow x,\quad\gamma\rightarrow y,\quad\left.\frac{\partial h}{\partial y}\right|_{y=\gamma}\rightarrow\frac{\partial h}{\partial y}=\frac{dx}{dy}.\]
 With these replacements, Eq. (\ref{eq:Snell-alt-2}) gets re-expressed
in form as \begin{equation}
\frac{dx}{dy}=\frac{y}{\beta-x-\frac{n_{\phi}}{n_{\theta}}\left[\left(\beta-x\right)^{2}+y^{2}\right]^{1/2}}.\label{eq:dxdy0}\end{equation}

How is the instantaneous focal function, $\beta,$ restricted? The
$\beta$ in Eq. (\ref{eq:dxdy0}) is restricted so that the expression
for $dx/dy$ does not blow up. Equation (\ref{eq:dxdy0}) is well
defined if and only if the denominator satisfies the condition, \begin{equation}
\beta-x-\frac{n_{\phi}}{n_{\theta}}\left[\left(\beta-x\right)^{2}+y^{2}\right]^{1/2}\neq0.\label{eq:b-x-start-comp}\end{equation}
 Contrarily, but equivalently, the previous statement can be reworded
as follows. Equation (\ref{eq:dxdy0}) is ill defined if and only
if the denominator satisfies the condition, \begin{equation}
\beta-x-\frac{n_{\phi}}{n_{\theta}}\left[\left(\beta-x\right)^{2}+y^{2}\right]^{1/2}=0.\label{eq:b-x-start}\end{equation}
 For reasons to follow, I shall proceed with the latter. To solve
for $\beta,$ I shall first rearrange Eq. (\ref{eq:b-x-start}) as
\begin{equation}
\beta-x=\frac{n_{\phi}}{n_{\theta}}\left[\left(\beta-x\right)^{2}+y^{2}\right]^{1/2}.\label{eq:b-x-first}\end{equation}
Squaring of both sides give \[
\left(\beta-x\right)^{2}=\frac{n_{\phi}^{2}}{n_{\theta}^{2}}\left(\beta-x\right)^{2}+\frac{n_{\phi}^{2}}{n_{\theta}^{2}}y^{2}.\]
 This expression can be rearranged to become \[
\left(\beta-x\right)^{2}\left(1-\frac{n_{\phi}^{2}}{n_{\theta}^{2}}\right)=\frac{n_{\phi}^{2}}{n_{\theta}^{2}}y^{2}\]
 or \[
\left(\beta-x\right)^{2}\left(\frac{n_{\theta}^{2}-n_{\phi}^{2}}{n_{\theta}^{2}}\right)=\frac{n_{\phi}^{2}}{n_{\theta}^{2}}y^{2}.\]
 And, solving for $\left(\beta-x\right),$ I obtain \begin{align}
\beta-x & =\pm\frac{n_{\phi}y}{\sqrt{n_{\theta}^{2}-n_{\phi}^{2}}},\label{eq:b-x-second}\end{align}
 where the $\pm$ came from the action of taking the  square root
on both sides, of course. Now, one of the signs in Eq. (\ref{eq:b-x-second})
can be eliminated by comparing with Eq. (\ref{eq:b-x-first}). This
is the reason why I proceeded with Eq. (\ref{eq:b-x-start}) instead
of Eq. (\ref{eq:b-x-start-comp}). The $n_{\phi}$ and $n_{\theta}$
in Eq. (\ref{eq:b-x-first}) are both real refractive indices, which
cannot be negative numbers. The instantaneous focal function, $\beta,$
and the lens thickness, $x,$ must be real, which implies $\left[\left(\beta-x\right)^{2}+y^{2}\right]$
must be non-negative else $\left[\left(\beta-x\right)^{2}+y^{2}\right]^{1/2}$
becomes an imaginary term. As real refractive indices cannot be negative
numbers, the term $\left(\beta-x\right)$ is also a non-negative real
in Eq. (\ref{eq:b-x-first}), provided $n_{\phi}>0,$ $n_{\theta}>0,$
and $\left[\left(\beta-x\right)^{2}+y^{2}\right]>0,$ of course. Therefore,
the right hand side of Eq. (\ref{eq:b-x-second}) must be positive;
and, this gives \[
\beta=x+\frac{n_{\phi}y}{\sqrt{n_{\theta}^{2}-n_{\phi}^{2}}}.\]
 Now, this is precisely the condition for $\beta$ which makes Eq.
(\ref{eq:dxdy0}) ill defined. Equivalently, then Eq. (\ref{eq:dxdy0})
becomes well behaved for $\beta$ satisfying the condition given by
\begin{equation}
\beta\neq x+\frac{n_{\phi}y}{\sqrt{n_{\theta}^{2}-n_{\phi}^{2}}}.\label{eq:beta-restriction}\end{equation}
 Equation (\ref{eq:beta-restriction}) defines the restriction for
the instantaneous focal function, $\beta.$ 

What can be concluded of the restriction so defined in Eq. (\ref{eq:beta-restriction})
for the instantaneous focal function? To answer this, recall that
terms such as $\beta,$ $x,$ $y,$ $n_{\phi},$ and $n_{\theta}$
are all real values.  And, there are no restrictions on $n_{\phi}$
and $n_{\theta}$ to speak of which of the two must be bigger or smaller
in value. Interesting per se, the choice of $n_{\phi}>n_{\theta}$
results in the statement, \begin{eqnarray}
\beta\neq x+\frac{in_{\phi}y}{\sqrt{n_{\phi}^{2}-n_{\theta}^{2}}}, &  & n_{\phi}>n_{\theta},\label{eq:beta-rest-test}\end{eqnarray}
 where the $i$ denotes the imaginary symbol and the term $\sqrt{n_{\theta}^{2}-n_{\phi}^{2}}$
in Eq. (\ref{eq:beta-restriction}) has been modified to $\sqrt{n_{\phi}^{2}-n_{\theta}^{2}}.$
But, this condition defined in Eq. (\ref{eq:beta-rest-test}) is always
satisfied, as $\beta$ is a real function. Therefore, it is concluded
that Eq. (\ref{eq:dxdy0}) is well behaved everywhere for $n_{\phi}>n_{\theta}.$

\section{Result}

\subsection{Lens surface equation}

The profile of axially symmetric lens about its optical axis is obtained
by solving the initial-value differential equation, Eq. (\ref{eq:dxdy0}),
\[
\frac{dx}{dy}=\frac{y}{\beta-x-\frac{n_{\phi}}{n_{\theta}}\left[\left(\beta-x\right)^{2}+y^{2}\right]^{1/2}},\quad x\left(y_{\textup{0}}\right)=x_{\textup{0}},\]
 where $x\left(y_{\textup{0}}\right)=x_{\textup{0}}$ is the initial
condition to be specified and the instantaneous focal function $\beta$
satisfies the constrain defined in Eq. (\ref{eq:beta-restriction}).
Without loss of generality, one may choose $x\left(y=y_{\textup{0}}=0\right)=0$
for the initial condition and the lens profile satisfies the differential
equation, \begin{align}
\frac{dx}{dy} & =\frac{y}{\beta-x-\frac{n_{\phi}}{n_{\theta}}\left[\left(\beta-x\right)^{2}+y^{2}\right]^{1/2}},\nonumber \\
\label{eq:ODE}\\ & x\left(0\right)=0,\quad\beta\neq x+\frac{n_{\phi}y}{\sqrt{n_{\theta}^{2}-n_{\phi}^{2}}}.\nonumber \end{align}
 The quantities $n_{\phi}$ and $n_{\theta}$ are the two refractive
indices in which one represents the lens and the other representing
the surrounding medium. Which of the two refractive indices corresponds
to the lens depends on the configuration of the problem, as demonstrated
in the proceeding  sections.

\subsection{Alexander and Lukyanov lens}

\subsubsection{Instantaneous focal function}

The instantaneous focal function proposed by Alexander and Lukyanov
has been discussed previously in Fig. \ref{fig:lens-data-Getman}.
The instantaneous focal function for each of the twelve zones can
be curve fitted and represented by a quadratic polynomial of the form
given by  \begin{eqnarray}
\beta\equiv\beta_{i}=ay_{i}^{2}+by_{i}+c, &  & y_{i,\textup{min}}\leq y_{i}\leq y_{i,\textup{max}},\label{eq:beta-i}\end{eqnarray}
 where the subscript $i$ of $\left(\beta_{i},\, y_{i},\, y_{i,\textup{min}},\, y_{i,\textup{max}}\right)$
denotes the $i\,\textup{th}$ concentric zone. The coefficients $a,$
$b,$ and $c,$ and the range for $y,$ which defines the width for
each of the axially symmetric concentric zones, are summarized in
Table \ref{tab:table1}. Since the instantaneous focal function, $\beta,$
and the lens radius, $y,$ have units of length measured in meters
$\left[\textup{m}\right],$ the coefficient $a$ must have a unit
of $\left[\textup{m}^{-1}\right],$ $c$ a unit of $\left[\textup{m}\right],$
and $b$ must be a unit-less scalar. To reduce the width of the table,
couple columns were represented in millimeter units, $\left[\textup{mm}\right].$ 

\begin{table}[H]
\caption{Domain $y_{i}$ and coefficients ($a,b,c$) of $\beta_{i}=ay_{i}^{2}+by_{i}+c$
for Fig. \ref{fig:lens-data-Getman} \label{tab:table1}}

\begin{centering}
\begin{tabular}{|c|c|c|c|}
\hline 
$y_{i,\textup{min}},\, y_{i,\textup{max}}\,$ {[}mm] & $a\,${[}1/m] & $b$ & $c\,${[}mm]\tabularnewline
\hline
\hline 
0.0, 0.19182692 & -313.07 & 0.0235 & 3.5137034\tabularnewline
\hline 
0.19519231, 0.27259615 & 534.53 & -0.2472 & 3.527877626\tabularnewline
\hline 
0.27596154, 0.33317308 & -309.02 & 0.0818 & 3.5088062\tabularnewline
\hline 
0.33653846, 0.38701923 & 536.05 & -0.3275 & 3.5493232\tabularnewline
\hline 
0.39038462, 0.43413462 & -306.12 & 0.1182 & 3.502912672\tabularnewline
\hline 
0.4375, 0.47451923 & 539.03 & -0.3891 & 3.569538239\tabularnewline
\hline 
0.47788462, 0.51153846 & -303.68 & 0.1463 & 3.496845208\tabularnewline
\hline 
0.51490385, 0.54855769 & 542.21 & -0.4417 & 3.589312176\tabularnewline
\hline 
0.55192308, 0.58221154 & -301.27 & 0.1695 & 3.49080193\tabularnewline
\hline 
0.58557692, 0.6125 & 545.96 & -0.4895 & 3.609151039\tabularnewline
\hline 
0.61586538, 0.64278846 & -298.81 & 0.1893 & 3.484870978\tabularnewline
\hline 
0.64615385, 0.67307692 & 179.08 & -0.0474 & 3.469596542\tabularnewline
\hline
\end{tabular}
\par\end{centering}
\end{table}

\subsubsection{Lens profile}

\begin{figure}[H]
\begin{centering}
\includegraphics[scale=0.65]{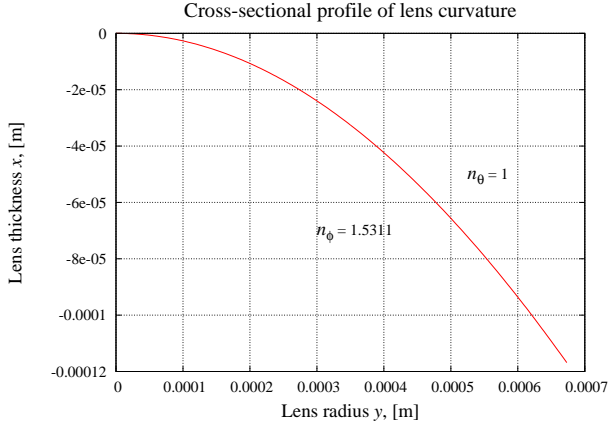}
\par\end{centering}

\caption{\label{fig:lens_cross_section} The cross-sectional profile of lens
curvature corresponding to the instantaneous focal function proposed
by Alexander and Lukyanov, Fig. \ref{fig:lens-data-Getman}. }

\end{figure}

\begin{figure}[H]
\begin{centering}
\includegraphics[scale=0.45]{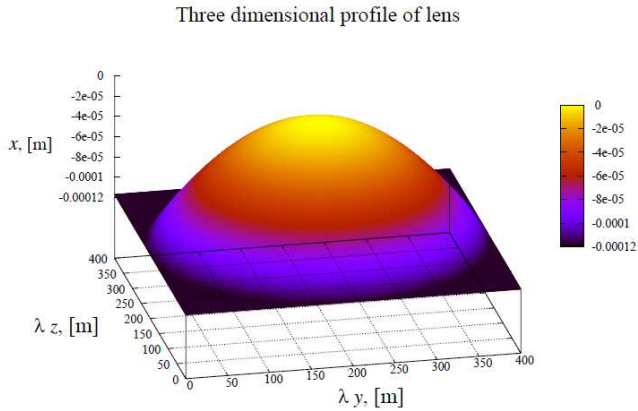}
\par\end{centering}

\caption{\label{fig:3D_lens_profile} Three dimensional profile of lens satisfying
the focal function $\beta$ proposed by Alexander and Lukyanov, Fig.
\ref{fig:lens-data-Getman}. The $y$ and $z$ axes have been multiplied
by $\lambda=3.365385\times10^{-6}.$ }

\end{figure}

Equation (\ref{eq:ODE}) was solved by the Runge-Kutta routine coded
in \textbf{FORTRAN 90}\citet{Derrick-Grossman}. For the computation,
refractive indices, $n_{\phi}$ and $n_{\theta},$ were chosen as
follows, \begin{eqnarray*}
n_{\phi}=1.5311, &  & n_{\theta}=1.0.\end{eqnarray*}
 In this configuration, $n_{\phi}$ denotes the refractive index for
the lens and $n_{\theta}$ denotes the refractive index for air.%
\footnote{Although the actual value varies depending on the surrounding humidity
and so on, the refractive index of an air is about $n\approx1.0008.$%
} The resulting cross-sectional profile of the lens curvature is shown
in Fig. \ref{fig:lens_cross_section}. The profile in Fig. \ref{fig:lens_cross_section}
was revolved about the optical axis, which is the $x$ axis in the
figure, to generate the three dimensional profile of the physical
lens. This result is shown in Fig. \ref{fig:3D_lens_profile}, where
the optical axis is located at $\left(\lambda y=200,\lambda z=200\right).$
The scaling factor of $\lambda=3.365385\times10^{-6}$ was introduced
for graphing purpose only. 

In spite of the non constant $\beta$ for the instantaneous focal
function (see Fig. \ref{fig:lens-data-Getman}), the resulting \textbf{EDoF}
lens shown in Figs. \ref{fig:lens_cross_section} and \ref{fig:3D_lens_profile}
seems to resemble the parabolic curve, which configuration is known
to have only one focal point, i.e., $\beta=\textup{constant}.$\citet{thomas-finney}.
Is the result portrayed in Fig. \ref{fig:lens_cross_section} (or
Fig. \ref{fig:3D_lens_profile}) correct? To give a qualitative answer
to this, I shall recall the \textbf{EDoF} lens parameter $l,$ which
was previously illustrated in Fig. \ref{fig:EDoF_lens}. 

In explicit form, the \textbf{EDoF} lens parameter $l$ is defined
as \begin{eqnarray}
l=\left\Vert \beta_{\textup{max}}-\beta_{\textup{min}}\right\Vert , &  & \left\{ \begin{array}{c}
\beta_{\textup{max}}>0,\\
\beta_{\textup{min}}>0,\end{array}\right.\label{eq:EDoF_parameter_L}\end{eqnarray}
 where $\beta_{\textup{max}}$ and $\beta_{\textup{min}}$ denote,
respectively, the maximum and the minimum values in the profile of
instantaneous focal function. In the focal function profile of Alexander
and Lukyanov, Fig. \ref{fig:lens-data-Getman}, $\beta_{\textup{max}}\approx3.5188\,\textup{mm}$
and $\beta_{\textup{min}}\approx3.4831\,\textup{mm}.$ Plugging the
information into Eq. (\ref{eq:EDoF_parameter_L}), this roughly gives
$l\approx36\,\textup{um}.$ The resemblance of the \textbf{EDoF} lens
to the parabolic curve can be attributed to the small value for $l.$
Considering that Alexander and Lukyanov's lens has a maximum radius
of $y\approx0.67\,\textup{mm},$ which can be identified from Table
\ref{tab:table1}, the lens diameter comes out to be about $d=2y\approx1.34\,\textup{mm}.$
This implies, the lens diameter is larger than the \textbf{EDoF} lens
parameter, $l,$ by a factor of thirty seven. Under such circumstance,
the lens could be perceived as having a single focal point from the
perspective of human eye. In spite of the existence of number of very
closely spaced focal points within the length of $l$ along the optical
axis, the human eyes do not have sufficient resolving power to distinguish
those focal points. Even less so, the human eyes cannot distinguish
the actual cross-sectional profile of the lens of which is only slightly
perturbed from the cross-sectional profile of the parabolic lens.
To prove that this is indeed the case, I shall use the derived lens
formula, Eq. (\ref{eq:ODE}), to generate various parabolic lenses.

\subsection{Validation of the result }

How does one know that Eq. (\ref{eq:ODE}) generates  the correct
profile for the lens? The easiest way to settle this dilemma is to
actually apply Eq. (\ref{eq:ODE}) to the well known types, i.e.,
the parabolic lenses.

\subsubsection{Simple parabolic lens}

\begin{figure}[H]
\begin{centering}
\includegraphics[scale=0.65]{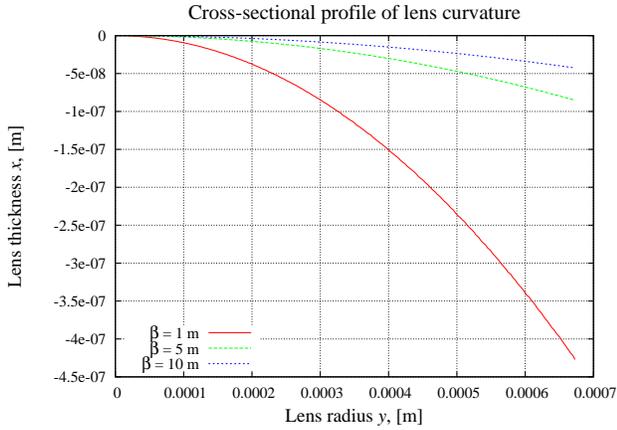}
\par\end{centering}

\caption{\label{fig:lens_cross_section_c1_c2_c3} The cross-sectional profile
of lens curvature for $\beta=1\,\textup{m},$ $\beta=5\,\textup{m},$
and $\beta=10\,\textup{m}.$}

\end{figure}

\begin{figure}[H]
\begin{centering}
\includegraphics[scale=0.45]{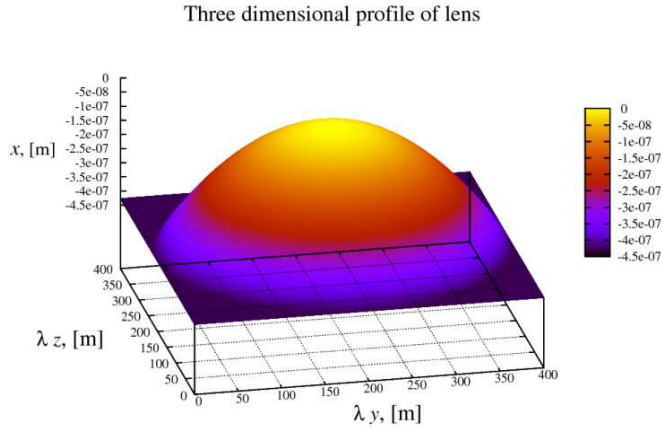}
\par\end{centering}

\caption{\label{fig:3D_lens_profile_c1} Three dimensional profile of a lens
with a constant instantaneous focal function,  $\beta=1\,\textup{m}.$
The $y$ and $z$ axes have been multiplied by $\lambda=3.365385\times10^{-6}.$ }

\end{figure}

The parabolic curves are known to merge parallel rays of incidence
light to a unique focal point called a focus\citet{thomas-finney}.
As a consequence of this, the parabolic curves flatten in the curvature
with the focus positioned at distances further from the vertex. Such
property of parabolic curves make them ideal for testing and validating
the lens formula defined in Eq. (\ref{eq:ODE}). The constant focus
of $\beta=1\,\textup{m},$ $\beta=5\,\textup{m},$ and $\beta=10\,\textup{m}$
were considered to generate curves using Eq. (\ref{eq:ODE}); and,
the result is summarized in Fig. \ref{fig:lens_cross_section_c1_c2_c3}.
 As expected, the generated curves are that of parabolic curves in
which each curves corresponds to focal points $\beta=1\,\textup{m},$
$\beta=5\,\textup{m},$ and $\beta=10\,\textup{m}.$ The curve corresponding
to $\beta=10\,\textup{m}$ is more flat in curvature than the ones
corresponding to $\beta=1\,\textup{m}$ or $\beta=5\,\textup{m},$
as expected. The result corresponding to $\beta=1\,\textup{m}$ was
revolved about the optical axis to illustrate the three dimensional
profile of the parabolic lens. This is shown in Fig. \ref{fig:3D_lens_profile_c1}.
Again, in the figure, the optical axis is  at $\left(\lambda y=200,\lambda z=200\right),$
where the scaling factor $\lambda$ is $\lambda=3.365385\times10^{-6}.$ 

\subsubsection{Concentric parabolic  lens}

\begin{figure}[H]
\begin{centering}
\includegraphics[scale=0.65]{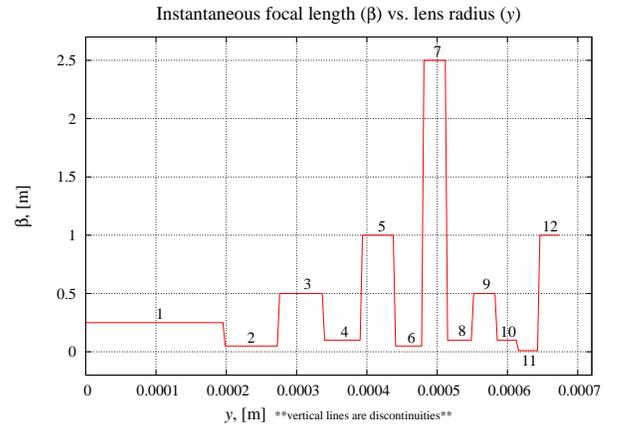}
\par\end{centering}

\caption{\label{fig:lens-data-0_test} Focal function $\beta$ with characteristic
of a step function. }

\end{figure}

\begin{figure}[H]
\begin{centering}
\includegraphics[scale=0.65]{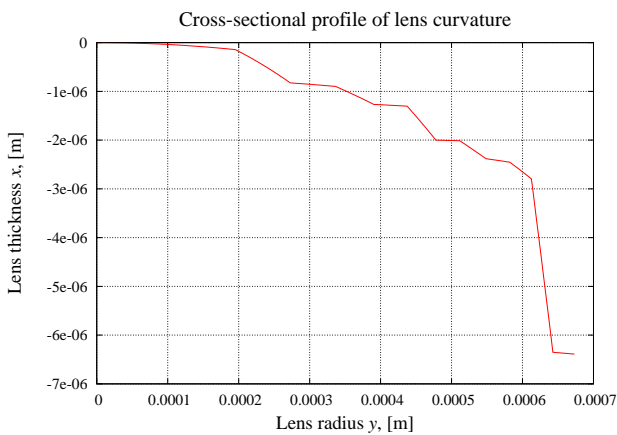}
\par\end{centering}

\caption{\label{fig:lens_cross_section_test} The cross-sectional profile of
lens curvature corresponding to the focal function defined in Fig.
\ref{fig:lens-data-0_test}. }

\end{figure}

\begin{figure}[H]
\begin{centering}
\includegraphics[scale=0.45]{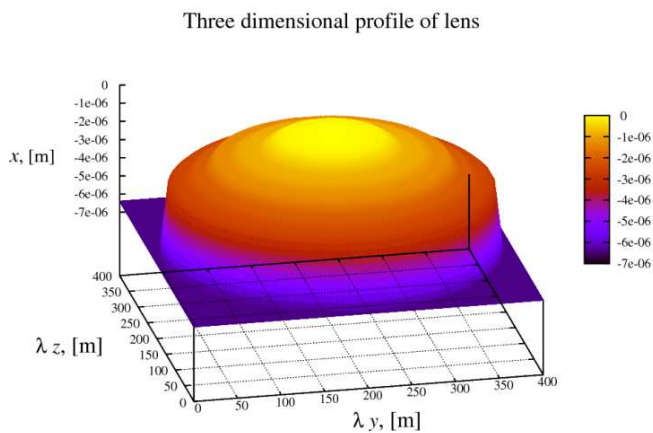}
\par\end{centering}

\caption{\label{fig:3D_lens_profile_test} Three dimensional profile of lens
satisfying the focal function defined in Fig. \ref{fig:lens-data-0_test}.
The $y$ and $z$ axes have been multiplied by $\lambda=3.365385\times10^{-6}.$ }

\end{figure}

Now I consider a slightly more complicated profile for the instantaneous
focal function, $\beta.$ To show that, indeed, the likeliness of
Alexander and Lukyanov's \textbf{EDoF} lens to the parabolic lens
is attributed to the small value for the \textbf{EDoF} lens parameter
$l,$ I shall modify only the $\beta$ portion of Alexander and Lukyanov's
profile for the instantaneous focal function, while leaving the size
of lens diameter unmodified. The modified focal function for this
test is shown in Fig. \ref{fig:lens-data-0_test}. And, the coefficients
$a,$ $b,$ and $c,$ and the range of $y$ corresponding to the $\beta\equiv\beta_{i}$
of Eq. (\ref{eq:beta-i}) for each of the twelve zones is summarized
in Table \ref{tab:table-test}. Besides the increased focal length
for each of twelve zones, the focal point is unique within each zones
in this test configuration. In Alexander and Lukyanov's focal profile,
$l$ was much smaller than the lens diameter $d,$ i.e., $d\approx37l.$
The $\beta_{\textup{max}}$ and $\beta_{\textup{min}}$ for the test
configuration are, respectively, $\beta_{\textup{max}}=2.5\,\textup{m}$
and $\beta_{\textup{min}}=0.01\,\textup{m},$ which can be verified
from Table \ref{tab:table-test}. Using the formula for $l$ defined
in Eq. (\ref{eq:EDoF_parameter_L}), this gives $l\approx2.49\,\textup{m}.$
Therefore, in this test configuration, $l$ is much larger than $d,$
i.e., $d\approx5.4\times10^{-4}l,$ which is just the opposite situation
from that of Alexander and Lukyanov. That being said, the Eq. (\ref{eq:ODE})
was plotted for the curve and the result is shown in Fig. \ref{fig:lens_cross_section_test}.
As expected, the resulting cross-sectional profile for the lens does
not resemble simple parabolic lens. However, the curve profile for
each of the twelve zones in Fig. \ref{fig:lens_cross_section_test}
represents the portion of a parabolic curve corresponding to $\beta_{i}$
illustrated in Fig. \ref{fig:lens-data-0_test}. By superimposing
the two graphs, Figs. \ref{fig:lens-data-0_test} and \ref{fig:lens_cross_section_test},
the boundaries for each zones can be identified by kinks in Fig. \ref{fig:lens_cross_section_test}.
The cross-sectional profile of the lens curvature illustrated in Fig.
\ref{fig:lens_cross_section_test} was revolved about the optical
axis for the three dimensional profile of the lens. This result is
shown in Fig. \ref{fig:3D_lens_profile_test}. 

\begin{table}[H]
\caption{Domain $y_{i}$ and coefficients ($a,b,c$) of $\beta_{i}=ay_{i}^{2}+by_{i}+c$
for Fig. \ref{fig:lens-data-0_test}\label{tab:table-test}}

\begin{centering}
\begin{tabular}{|c|c|c|c|}
\hline 
$y_{i,\textup{min}},\, y_{i,\textup{max}}\,$ {[}mm] & $a\,${[}1/m] & $b$ & $c\,${[}m]\tabularnewline
\hline
\hline 
0.0, 0.19182692 & 0.0 & 0.0 & 0.25\tabularnewline
\hline 
0.19519231, 0.27259615 & 0.0 & 0.0 & 0.05\tabularnewline
\hline 
0.27596154, 0.33317308 & 0.0 & 0.0 & 0.50\tabularnewline
\hline 
0.33653846, 0.38701923 & 0.0 & 0.0 & 0.10\tabularnewline
\hline 
0.39038462, 0.43413462 & 0.0 & 0.0 & 1.00\tabularnewline
\hline 
0.4375, 0.47451923 & 0.0 & 0.0 & 0.05\tabularnewline
\hline 
0.47788462, 0.51153846 & 0.0 & 0.0 & 2.50\tabularnewline
\hline 
0.51490385, 0.54855769 & 0.0 & 0.0 & 0.10\tabularnewline
\hline 
0.55192308, 0.58221154 & 0.0 & 0.0 & 0.50\tabularnewline
\hline 
0.58557692, 0.6125 & 0.0 & 0.0 & 0.10\tabularnewline
\hline 
0.61586538, 0.64278846 & 0.0 & 0.0 & 0.01\tabularnewline
\hline 
0.64615385, 0.67307692 & 0.0 & 0.0 & 1.00\tabularnewline
\hline
\end{tabular}
\par\end{centering}
\end{table}

Basing on these results, it can be concluded that for the case where
the lens diameter $d$ is much larger than the \textbf{EDoF} lens
parameter $l,$ the profile of the lens resembles closely the profile
of simple parabolic lens. Here, the word {}``simple'' has been used
to denote the parabolic curve with single focus. In the opposite situation,
where the lens diameter $d$ is much smaller than the \textbf{EDoF}
lens parameter $l,$ the profile of the lens no longer resembles the
simple parabolic lens. Instead, in this latter case, the shape for
the lens resembles superimposed, multiple number of parabolic lenses
of different degrees of curvature, as illustrated in Fig. \ref{fig:3D_lens_profile_test}.

\section{Concluding Remarks}

The image processing speed and the quality of processed images are
both of critical importance in software assisted imaging technology.
Such a requirement calls for the optimization of image reconstruction
code based on the principles of wave optics. The coding side of the
\textbf{SAIT} system can be optimized if \textbf{EDoF} lens is used
for the input stage. 

In this presentation, the formula for the \textbf{EDoF} lens has been
derived based on the knowledge of instantaneous focal function, $\beta.$
The $\beta$ information is an important tool in the design of \textbf{EDoF}
lens, as this allows  optical engineer to try out various mathematical
curves using computers for optimization. With the knowledge of $\beta,$
this can be achieved without having to actually make \textbf{EDoF}
lens prototypes, thereby saving time and the cost. Once the optimal
solution for the instantaneous focal function is obtained, the physical
\textbf{EDoF} lens can be manufactured based on the lens formula presented
in this work.

\section{Acknowledgments}

I would like to thank G. Alexander for  providing the raw data for
the focal function. I would also like to thank Dr. S. Lee for generating
the \textbf{PSF} information for the lens designed in this work using
CODE V\textregistered{}. The author acknowledges the support for this
work provided by Samsung Electronics, Ltd.

\end{document}